\documentclass[aps,prl,reprint,groupedaddress]{revtex4-2}
\usepackage{graphicx}% Include figure files
\usepackage{dcolumn}% Align table columns on decimal point
\usepackage{bm}% bold math
\usepackage[mathlines]{lineno}% Enable numbering of text and display math
\usepackage{comment}
\usepackage{amsmath}
\usepackage{amssymb}
\usepackage{mathrsfs}
\usepackage{tensor}
\usepackage{booktabs}
\usepackage{multirow}
\usepackage{xcolor}
\usepackage{subfigure}
\usepackage{braket}
\usepackage{siunitx}
\usepackage{longtable}
\usepackage{etoolbox}
\usepackage{soul}
\usepackage[colorlinks]{hyperref}% add hypertext capabilities
\usepackage{ulem}
 \hypersetup{
     colorlinks=true,
     linkcolor=blue,
     filecolor=blue,
     citecolor = blue,
     urlcolor=blue,
 }
\newcommand{\be}{\begin{equation}}
\newcommand{\ee}{\end{equation}}
\newcommand{\bea}{\begin{eqnarray}}
\newcommand{\eea}{\end{eqnarray}}
\newcommand{\lan}{\langle}
\newcommand{\ran}{\rangle}

\newcommand{\bn}{{\bar{\nu}}}

\newcommand{\nue}{\nu_e}
\newcommand{\numu}{\nu_\mu}
\newcommand{\nuta}{\nu_\tau}

\newcommand{\ga}{\gamma}

\begin{document}

%\preprint{}

\title{Lepton Double Charge Exchange Reactions as Probes for Lepton Number Violation}

\author{H. Lenske}
\affiliation{Institut f\"ur Theoretische Physik, Justus-Liebig-Universit\"at Giessen, 35392 Gießen, Germany}
%\thanks{}

\author{F. Cappuzzello}
\affiliation{Dipartimento di Fisica e Astronomia "Ettore Majorana", Universit\`a di Catania, Catania, Italy}
\affiliation{Istituto Nazionale di Fisica Nucleare,  Laboratori Nazionali del Sud, Catania, Italy}

\author{A. Spatafora}
\affiliation{Istituto Nazionale di Fisica Nucleare,  Laboratori Nazionali del Sud, Catania, Italy}

%\begin{document}

\date{\today}

\begin{abstract}
Lepton number violating (LNV) $A(e^-,e^+)X$ double charge exchange (LDCE) reactions on nuclei at accelerator facilities with multi-GeV beams are proposed as a probe for physics beyond the Standard Model (BSM). A second order formalism is presented including LNV dynamics by the left-right symmetric model (LRSM). For practical studies a phenomenological model is used to estimate LDCE cross sections numerically. Sizable cross sections are predicted for multi-GeV beam energies. LDCE reactions proceed preferentially by the energy-momentum dependent left-right mixing terms. While Majorana mass terms are negligible for light neutrinos, they may become sizable for heavy neutral leptons at higher beam energies.
In the 10~GeV region inclusive total LDCE cross sections of about $100\times|\Gamma_{BSM}|^2$~fb in units of the BSM vertices are predicted, increasing strongly with energy and target mass. LDCE experiments seem to be feasible with existing equipment under full laboratory control and free of the constraints imposed on decay or capture experiments of nuclear and hadronic systems.
\end{abstract}

%\keywords{}

\maketitle

\paragraph{\textbf{Introduction:}}
In the search for signatures of physics Beyond the Standard Model (BSM) lepton number violation (LNV) is probably the most promising candidate. Research is being pursued by neutrinoless double beta decay (DBD) at nuclear energies, LNV decays of hadrons, and tracing same-charge lepton pairs plus hadron di-jets (LLjj) events at the LHC and envisioned as a major activity for future collider experiments. Strategies how to prepare experimental conditions favoring or enhancing LNV events have been explored actively over decades \cite{PhysRevD.38.2102,Vergados:1981bm,Vergados:2002pv,Divari:2002sq,Missimer:1994xd,SINDRUMII:1991lmr,SINDRUMII:2006dvw,vanderSchaaf:2021hnd} and are the subject of currently intensified searches for alternative routes to BSM physics, e.g. \cite{Babu:2022ycv,Mikulenko:2023ezx,Yang:2025jxc,Liao:2016hru,Barry:2013xxa,Fridell:2023rtr,Antusch:2023jsa}.

\begin{figure}
	\centering
	\includegraphics[width=1.\linewidth]{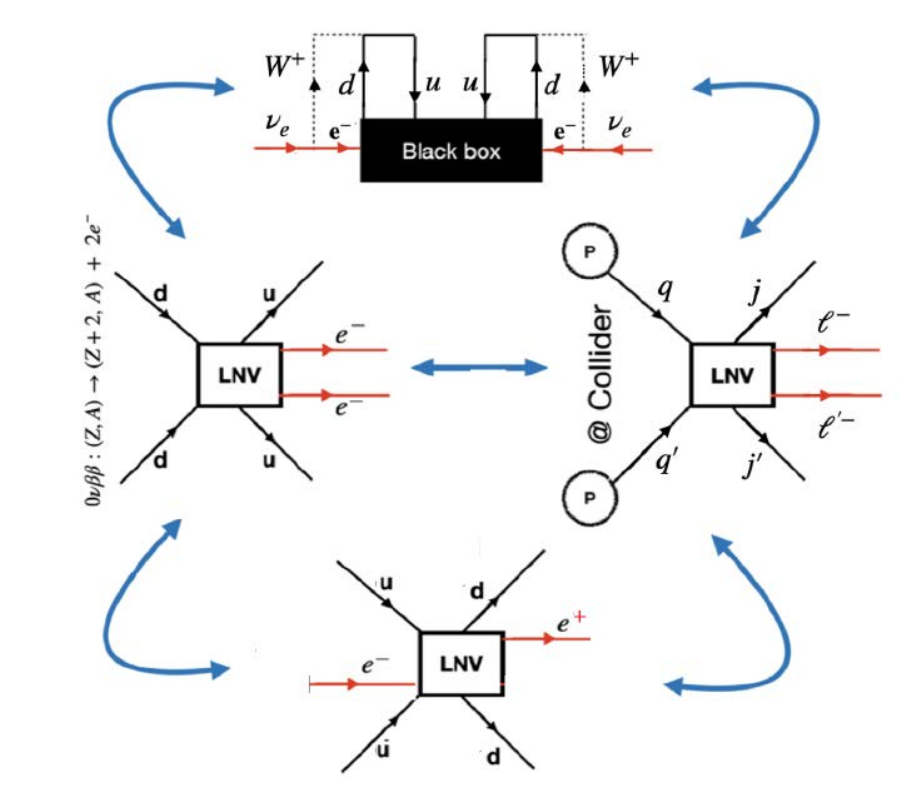}
	\caption{Pictorial illustration of the Black Box theorem \cite{Schechter:1982dbd} and its application to three physical processes, namely $0\nu\beta\beta$, $l^\pm l^\pm jj$ and LDCE, compliant with it.}
	\label{fig:BBT}
\end{figure}

The extreme tension between nuclear MeV-scales and the multi-TeV energies required for LLjj-events demands for methods being able to bridge that gap. The purpose of the present work is to point to the surprisingly unexplored potential of LNV investigations at accelerator facilities. As versatile LNV probes we propose ($e^-,e^+$) Lepton Double Charge Exchange (LDCE) reactions on nuclear targets. LDCE reactions at multi-GeV beam energies are free of the DBD nuclear structure constraints and are feasible with largely available technology. Although such reactions are centered in the non-perturbative regime, they will be extremely useful to gain insight into LNV dynamics, not to the least as preparation for the much more involved LLjj measurements.

A convincing way to link conceptually diverse LNV processes is offered by the so-called Black Box Theorem (BBT), originally formulated by Schechter and Valle \cite{Schechter:1982dbd} and recently extended to $p+p\rightarrow l^\pm l^\pm jj$ reactions by Babu et al. \cite{Babu:2022ycv}. The BBT states that whatever mechanism is at the origin of $0\nu\beta\beta$ or high-energy LNV-dijet emission, the observation of one of such events will prove that (massive) neutrinos are Majorana fermions, $\nu_M=\nu^C_M=\bar{\nu}_M$. The generality of the theorem allows to include LDCE reactions and we may state that if one of the three processes exists also the others do. In Fig.~\ref{fig:BBT} the generalized BBT is illustrated pictorially.

\begin{figure}
	\centering
	\includegraphics[width=\linewidth]{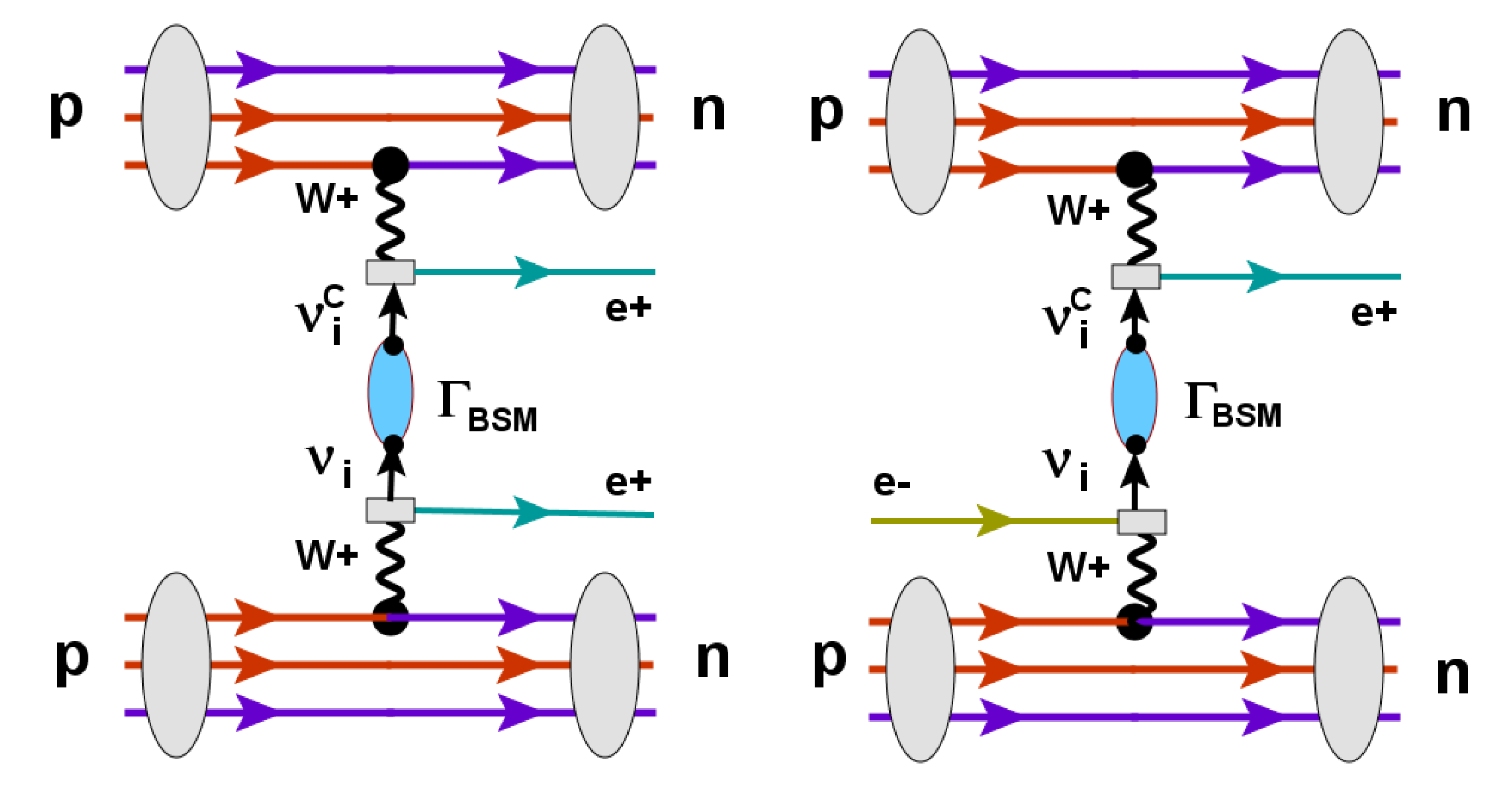}
	\caption{Nuclear $0\nu\beta\beta$ (left) and the quasi-elastic  $(e^-,e^+ )$ LDCE reaction (right), both inducing a $\Delta Z=-2$ nuclear transition through t-channel exchange of massive Majorana neutrinos $\nu_i=\nu^C_i$, are compared on the quark level. The LNV vertices (filled blue ellipsoids) are denoted by $\Gamma_{BSM}$. The SM-CC vertices are indicated by filled boxes. Time is running from left to right.}
	\label{fig:Feynman_diagrams}
\end{figure}

\paragraph{\textbf{LDCE Concepts:}}
LDCE reactions are second order double charged current (DCC) processes. In Fig.~\ref{fig:Feynman_diagrams} the elementary DBD and DCC processes are depicted on the lepton-quark level, showing their connection by crossing symmetry. However, neither DBD nor LDCE will directly probe the shown sub-nuclear weak interactions. Rather, they both correspond to a faint low-energy echo of the hard elementary LNV dynamics - which is the regime of LLjj physics.

The Left-Right Symmetric Model (LRSM)  provides the suitable tools for LNV studies by treating left-handed (LH) and right-handed (RH) lepton currents on equal footing, see e.g. \cite{Altarelli:2003ph,Buchmuller:2004nz,Kayser:2008rd,Barry:2013xxa}. Eliminating the gauge bosons in favor of their nucleonic sources the effective low-energy LRSM Lagrangian with lepton-nucleon current-current interactions is obtained:
\be\label{eq:Lcc}
\mathcal{L}_{CC}=\widetilde{G}_F
\left(j_{\mu L}(J^\mu_L+\kappa J^\mu_R)+j_{\mu R}(\lambda J^\mu_R+\eta J^\mu_L)
\right),
\ee
The strength
$\widetilde{G}_F=G_F\cos{\theta_C}/{\sqrt{2}}$ is defined by the Fermi constant $G_F$ and the  Cabibbo-Kobayashi-Maskawa (CKM) quark mixing angle $\theta_C$. 
Helicity-dependent lepton CC currents  $j_{\mu,L/R}=\bar{\ell}\ga_\mu (1-\gamma_5)\nu_{L/R}$ are coupled to nucleon isovector currents $J^\mu_{L/R}=J^\mu\mp J^\mu_5$. The latter are composed of vector and rank-2 tensor operators, contained in $J^\mu$, and axial vector and pseudo-scalar operators, included in $J^\mu_5$, respectively. The coefficients $\eta$, $\kappa$ account for LR-mixing induced by the presence of RH currents while $\lambda$ originates from LR mixing of the weak gauge bosons  \cite{Doi:1982dn,Tomoda:1990rs,Altarelli:2003ph,Buchmuller:2004nz,Kayser:2008rd,Barry:2013xxa,Vergados:2012xy}.

While charged leptons are flavor and mass eigenstates, the neutral current leptons $\{\nue,\numu,\nuta \}$ are flavor eigenstates only. By the Pontecorvo–Maki–Nakagawa–Sakata-(PMNS)-type \cite{Pontecorvo:1957qd,Maki:1960ut} matrices:
\bea\label{eq:PMNS}
|\nu_{eh}\ran= \sum_{i=1}^{N_M}U^{(h)}_{ei}|n_{ih}\ran \quad ; \quad
|\bar{\nu}_{eh}\ran = \sum_{i=1}^{N_M}V^{(h)}_{ei}|n_{ih}\ran ,
\eea
the flavor states of helicity $h=L,R$ are expressed as superpositions of the $N_M \geq 3$ mass eigenstates $|n_{ih}\ran$ of masses $m_i$.
The summations include the full set of light (LNL) and expected heavy (HNL) massive neutral leptons.

\begin{figure}
	\centering
   \includegraphics[width=0.65\linewidth]{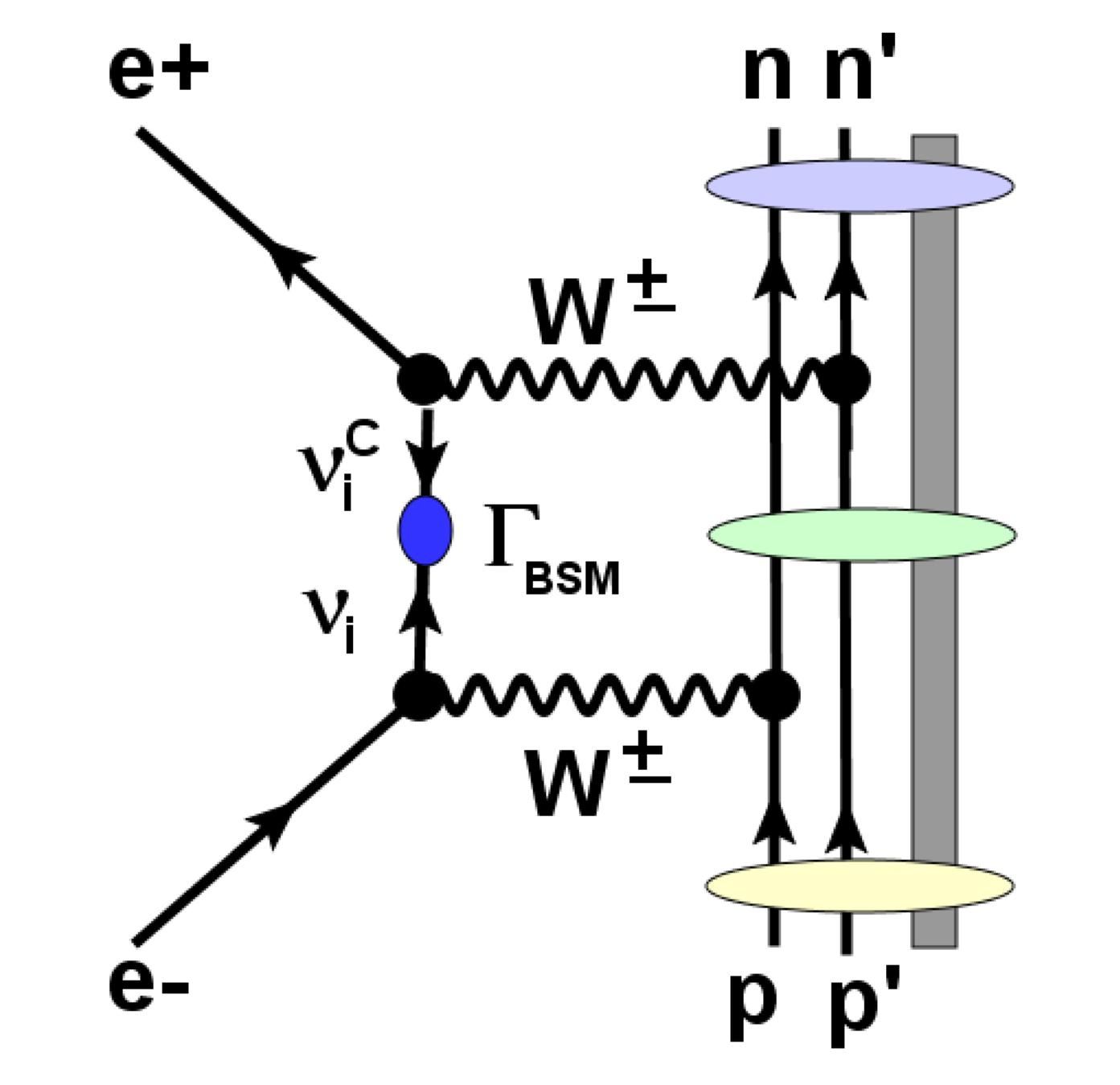}
	\caption{
    Feynman graph emphasizing the s-channel nature of a LDCE reaction $A\to B$ with a change of hadronic charge by $\Delta Z=-2$, illustrated for QE modes. The reaction proceeds either by virtual emission of $W^+$ bosons from the nucleus or by radiating $W^-$ bosons off weak lepton CC processes. The time order of the protons may be exchanged and short-range correlations will play a role. In practice, gauge boson propagation is replaced by contact interactions of strength $G_F\cos(\theta_C)/\sqrt{2}$ - see text. }
	\label{fig:Feynman_nucleus}
\end{figure}

As seen in Fig.~\ref{fig:Feynman_diagrams}, DBD-LNV is an effect of t-channel exchange of Majorana neutrinos which largely protects the process from outside interferences and inhibits manipulations of nuclear DBD-LNV events by experimental means.
A major motivation of LDCE physics is to overcome that limitation by exploiting the complementary s-channel LNV process allowing direct experimental access. In Fig.~\ref{fig:Feynman_nucleus} the proper definition of the second order LDCE reaction amplitude is displayed, illustrated for a quasi-elastic (QE) process. The process is initialized and terminated by SM ($e^-,\nue$) and ($\bn,e^+$) half off-shell CC transitions, respectively. The physical in- and out-channels are connected by the intermediate (off-shell) $n_i + C$ systems. In principle the entrance and exit matrix elements could be investigated independently by the respective on-shell CC reactions. Also neutrons will contribute to a LDCE reaction, e.g. by virtual or on-shell $n\to W^+ +\Delta^{-}$ and  $W^{-}+n\to \Delta^{-}$ processes or any other nucleon resonance or deep-inelastic scattering quark-configurations. Hence, LDCE reactions will increase $\sim \mathcal{O}(A^2)$.

\begin{figure}
	\centering
   \includegraphics[width=1.\linewidth]{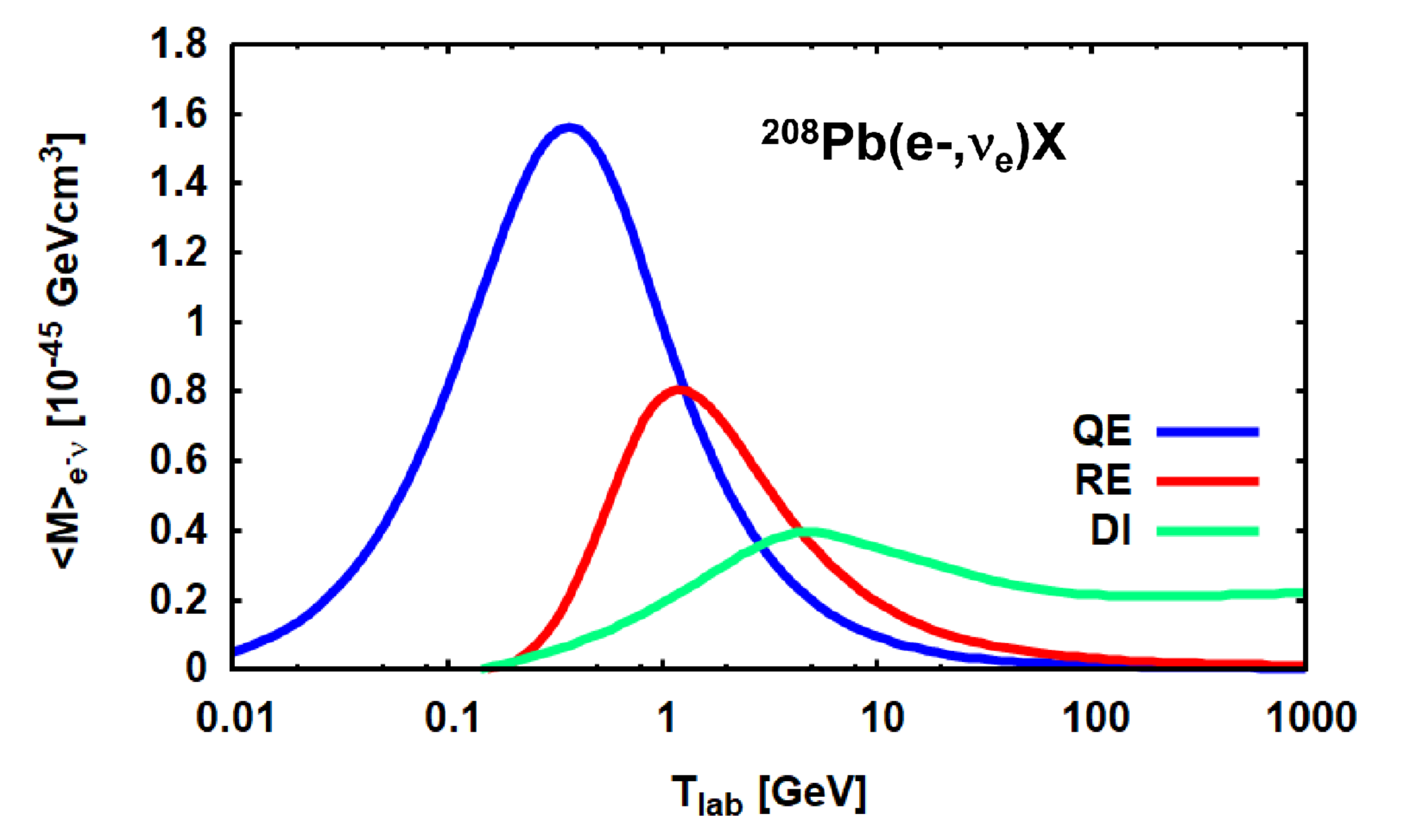}
   \includegraphics[width=1.\linewidth]{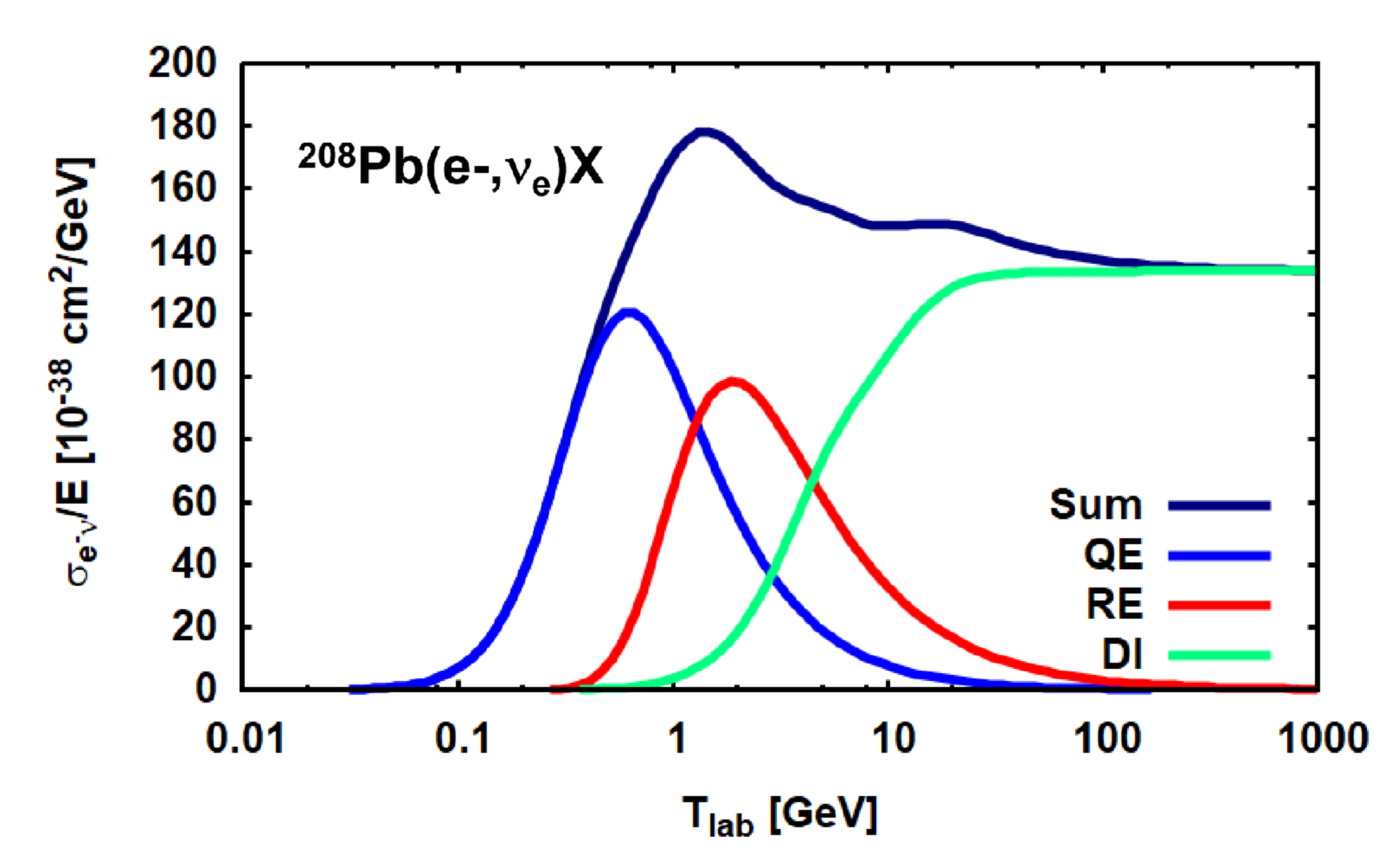}
	\caption{Charged current matrix elements (top) and total cross sections per nucleon and energy (bottom) for the
     $^{208}$Pb$(e^-,\nue)$X CC reaction. The QE, RE, and DI contributions as predicted by our
    phenomenological approach are also displayed. Statistical and systematical uncertainties of at least 20\% should be added.}
	\label{fig:enu}
\end{figure}

The $\alpha=e^- +A$ entrance channel defines the conserved baryon number $B_\alpha=A$ and charge number $Z_\alpha=Z_A-1$. The intermediate nuclear system $C$ is constrained by $Z_\alpha$ and $B_\alpha$, allowing, however, an arbitrary number of pions and photons. Similarly, the final hadronic system $X$ is a collection of configurations $X_f$ defined by charge and baryon numbers $Z_\alpha-1=Z_A-2$ and $B_\alpha$, respectively. The intermediate and final states are spread over a large spectrum of hadron energies, $0\leq E_x \leq T_\alpha$, where $T_\alpha=(s_\alpha-(m_e+M_A)^2)/2M_A$ is the invariant kinetic energy brought in by the electron beam. Hence, at multi-GeV beam-energies the spectra of $C$ and $X$ will
be probed from quasi-elastic (QE) nuclear modes over nucleon resonance excitations (RE) to deep-inelastic (DI) scattering, see Fig.~\ref{fig:enu} and \cite{Formaggio:2012cpf}.

\paragraph{\textbf{LDCE Reactions in the Left-Right Symmetric Model:}}
In operator form the LDCE reaction amplitude is readily written down as a matrix element of a second order two-body LNV operator playing the role of a field-theoretical generalization of the DBD neutrino potential: 
\be 
\mathcal{M}^{(e^-e^+)}_{AX_f}=
\lan e^+X_f|\mathcal{L}_{CC}\mathcal{G}\mathcal{L}_{CC}|Ae^-\ran.
\ee
Here, we retain the full second order structure and express the operator in terms of  helicity-projected lepton states $h_i$:
\be\label{eq:Mee}
\mathcal{M}^{(e^-e^+)}_{AX_f}=\sum_{h_1,h_2}\Gamma_{h_1h_2}\mathcal{M}^{(e^-h_1,e^+h_2)}_{AX_f}.
\ee
Eq.\eqref{eq:Lcc} shows that $\mathcal{M}^{(e^-e^+)}_{AX_f}$ is in fact a sum over $4 \times 4=16$ terms of second order amplitudes. However, only the seven leading order amplitudes, linear in $\{1,\eta,\kappa,\lambda\}$, are expected to be of significant strength.

LR-mixing is described by the BSM vertices $\Gamma_{h_1h_2}$, given by products of the LRSM coefficients $1,\eta,\kappa,\lambda$. The reaction is initialized and terminated by the SM-CC amplitudes 
$\mathcal{M}^{(e^-\nue,h_1)}_{AC}=\widetilde{G}_F\lan \nue,C|j_{h_1,\mu}J^\mu_{h'_1}|A,e^-\ran$ and  
$\mathcal{M}^{(\bn_e e^+,h_2)}_{CX_f}= \widetilde{G}_F\lan e^+,X_f|j_{h_2,\mu}J^\mu_{h'_2}|C,\bn_e\ran$, respectively. The complementary helicity status of the nuclear currents, see Eq.\eqref{eq:Lcc}, is denoted by $h'_{1,2}\in \{L,R\}$ and is included implicitly into the lepton $h_i$ summations.

In the basis of intermediate hadronic states $\{C\}$ the reaction matrix elements (RME) are
\bea
\mathcal{M}^{(e^-h_1,e^+h_2)}_{AX_f}=
\sum_C
\mathcal{M}^{(\bn_{e}e^+,h_2)}_{CX_f}
\mathcal{G}^{(h_1h_2)}_{\bn_e\nue}
\mathcal{M}^{(e^-\nue,h_1)}_{AC}.
\eea

The intermediate evolution of the system is described by the effective $\nue,\bn_e$ propagators
\be
\mathcal{G}^{(h_2h_1)}_{\bn_e\nue}=\sum_{i=1}^{N_M} V^{(h_2)}_{\bn_e n_i}
 P_{h_2}\mathcal{G}_{i C}P_{h_1} U^{(h_1\dag)}_{n_i\nue},
\ee
where  Eq.\eqref{eq:PMNS} was used and  $P_{L/R}=\frac{1}{2}(1 \mp \ga_5)$ denote the helicity projectors.

Reverting to momentum representation, the energy integration is performed first. In the $e^-A$ center-of-mass frame with  $P_\alpha=p_{e^-}+p_A=(w_\alpha,\mathbf{0})^T$ the result is
\bea \label{eq:MGM}
&&\mathcal{M}^{(e^-h_1,e^+h_2)}_{AX_f}(\mathbf{k}_\alpha,\mathbf{k}_\beta)=\\
&&\sum_C\int \frac{d^3k}{(2\pi)^3} \frac{M_C}{2E_C(\mathbf{k})}
\mathcal{M}^{(\bn_{e}e^+,h_2)}_{CX_f}(\mathbf{k}_\beta,\mathbf{k})\nonumber\\
&&\times
\left((1-\delta_{h_1h_2})\mathcal{D}^{(h_1,h_2)}_\mu(\mathbf{k})\ga^\mu+
\delta_{h_1h_2}\mathcal{S}^{(h_1,h_2)}(\mathbf{k}) \right)\nonumber\\
&&\times \mathcal{M}^{(e^-\nu_{e},h_1)}_{AC}(\mathbf{k},\mathbf{k}_\alpha)
.\nonumber
\eea
Helicity-diagonal scalar mass terms 
\be 
\mathcal{S}^{(h_1,h_2)}(\mathbf{k})= \sum_i V^{(h_2)}_{\bn_e i}m_i \frac{w_\alpha}{E_i(\mathbf{k})} D_2(\mathbf{k}) 
U^{(h_1)\dag}_{\nue i} 
\ee
and the helicity-non-diagonal four-vectors
\bea 
&&D^{(h_1,h_2)}_\mu(\mathbf{k})=\\
&&\sum_i V^{(h_2)}_{\bn_e i}
(E_i(\mathbf{k})D_1(\mathbf{k}),\mathbf{k}\frac{w_\alpha}{E_i(\mathbf{k})}D_2(\mathbf{k}))^TU^{(h_1)\dag}_{\nue i} \nonumber
\eea
are obtained. Assuming unitarity for the PMNS matrices, in first approximation $VU^\dag \approx 1$ may be used. 
The recoil-propagators
\be
D_{1,2}(\mathbf{k})=D^{(-)}_{1,2}(\mathbf{k})-D^{(+)}_{1,2}(\mathbf{k}),
\ee
are given by
\bea
&&D^{(\pm)}_1(\mathbf{k})=\frac{1}{(w_\alpha\pm E_C(\mathbf{k}))^2-E^2_i(\mathbf{k})+i\epsilon}\\
&&D^{(\pm)}_2(\mathbf{k})=\frac{1}{w^2_\alpha-(E_C(\mathbf{k})\pm E_i(\mathbf{k}))^2+i\epsilon}.
\eea
with poles at the on-shell three-momenta $\mathbf{k}_{iC}$, determined by 
$w_\alpha-E_C(\mathbf{k}_{iC})-E_i(\mathbf{k}_{iC})=0$.
LR-mixing effects are encoded also in t-channel DBD dynamics \cite{Tomoda:1990rs,Ejiri:2025oro}, but magnified significantly in LDCE reactions by the completely different s-channel energy dependencies.

The LDCE-RME are rearranged into total vector and  scalar amplitudes:
\be \label{eq:MeeVS}
\mathcal{M}^{(e^-e^+)}_{AX_f}=\ga_\mu \mathcal{V}^\mu_{AX_f}+ \mathcal{S}_{AX_f}
\ee
each given as a sum over helicity amplitudes, e.g.
\be
\mathcal{V}^\mu_{AX_f}=
\sum_{h_1,h_2}\Gamma_{h_1h_2}\mathcal{V}^\mu_{h_1h_2}.
\ee

\paragraph{\textbf{LDCE Cross Sections:}}
From Eq.\eqref{eq:MeeVS} we find
\bea \label{eq:MeeSquared}
|\mathcal{M}^{(e^-e^+)}_{AX_f}|^2&=&
\left(g^\nu_\mu+\frac{i}{2}\sigma_{\mu\nu} \right)\mathcal{V}^{\mu}_{AX_f}\mathcal{V}^{\nu\dag}_{AX_f}+|\mathcal{S}_{AX_f}|^2 \nonumber\\
&+&
2\ga_\mu\Re\left(\mathcal{V}^\mu_{AX_f}\mathcal{S}^\dag_{AX_f} \right).
\eea 
Upon evaluating traces over the $e^\mp$  spins, the relativistic vector and rank-2 tensor terms will vanish and we are left the spin-averaged amplitude $|\overline{\mathcal{M}}^{(e^-e^+)}_{AX_f}|^2=|\mathcal{V}_{AX_f}|^2+|\mathcal{S}_{AX_f}|^2$. Thus, without polarization
the LDCE double differential cross section in the center-of-mass frame is
\bea \label{eq:dsigma}
&&d^2\sigma^{(e^-e^+)}_{AX_f}(k_\alpha,\mathbf{k}_\beta)=\\
&&S(M_{X_f})\Pi_{AX_f} \sum_{[s]}|\overline{\mathcal{M}}^{(e^-e^+)}_{AX_f}(\mathbf{k}_\alpha,\mathbf{k}_\beta)|^2
dM_{X_f}d\hat{\mathbf{k}}_\beta,\nonumber
\eea
where $\mathbf{k}_{\beta}$ are the invariant three-momenta in the $e^+ + X_f$ exit channels. $S(M_{X_f })$ is the spectral distribution of final states $X_f$. The summation over spin quantum numbers of initial and final states is denoted by $[s]$ and a phase-space factor was introduced:
\bea
\Pi_{A X}=
\frac{1}{(2s_i+1)(2J_A+1)}
\frac{E_{e^-}(k_\alpha)E_{e^+}(k_\beta)}{(2\pi \hbar^2)^3w_\alpha}.
\eea
The cross section describes an inclusive reaction, integrated over the out-channels containing the prompt positron. According to Eq.\eqref{eq:Mee} the differential cross section is an incoherent sum of vector and scalar terms where each of them is a coherent superposition of interfering mixed-helicity amplitudes from intermediate massive Majorana neutrinos.

The total yield to be expected in a $A(e^-,e^+)X$ reaction is recorded by the
total LDCE cross section:
\bea \label{eq:XSee}
\sigma^{(e^-e^+)}_{A}(w_\alpha)&=&\int dM_{X_f}S(M_{X_f})\Pi_{AX_f}(k_\alpha,k_\beta)\\
&\times&\int d\hat{\mathbf{k}}_{\beta}\sum_{[s]}
|\overline{\mathcal{M}}^{(e^-e^+)}_{AX_f}(\mathbf{k}_\alpha,\mathbf{k}_\beta)|^2,\nonumber
\eea
To the extent that interference effects are averaging out, the total cross sections are sums of LR contributions weighted by the BSM factors
$|\Gamma_{BSM}|^2\in \{|\Gamma_{h_1h_2}|^2\}$, $h_{1,2}=L,R$.

A closer inspection reveals that the LR mixing terms will be significantly enhanced with increasing beam energy. The Majorana mass terms, however, will be quenched relative to the four-momentum mixing parts by factors $R_i=m_i/k$. At a beam energy of 10~GeV for a LNL of $m_i=1$~eV the quenching amounts to $R_i\approx 1.3\times 10^{-13}$ in the average for targets from $^{12}$C to $^{208}$Pb at the respective on-shell momenta $k_{iC}$. However, for HNLs of $m_i\simeq 100$~GeV one finds $R_i\sim \mathcal{O}(1)$ up to $A\sim 40$. For heavier targets and 10~GeV beams the respective channels move below threshold.

An interesting property of the LDCE amplitudes is revealed by studying the propagator spectrum. For negligibly small masses, $m_i\ll M_C$, the spectrum evolves $\sim T_{beam}+M_A-M_C=T_{beam}-Q_{AC}$, thus reflecting the spectrum of the hadronic configurations $C$. If, however, heavy neutrinos with $m_i\gtrsim M_C$ should exist, a notable shift of the pole spectrum $\sim T_{beam}-Q_{AC}-m_i$ towards lower intermediate energies will occur (and disappear below threshold if $m_i>T_{beam}-Q_{AC}$). That extra-shift will induce eventually an unexpected enhancement of LDCE cross sections already at lower beam energies.

\paragraph{\textbf{Estimating Incident Energy and Target Mass Dependencies:}} While CC reactions have been studied for decades in competing approaches up to about 2~GeV \cite{SajjadAthar:2022pjt,Alvarez-Ruso:2025oak}, the region up to and above incident energies of 10~GeV is largely \textit{terra incognita}. As to escape that dilemma we revert to a semi-phenomenological approach but keeping in mind the inherently large uncertainties. Here, the intention is to estimate realistically total LDCE cross sections.     

First,
the available $\nu_\mu$ and $\bar{\nu}_\mu$ CC cross section data \cite{Formaggio:2012cpf} are parameterized in functional form by Lorentzian and Gaussian distributions. Then, mean CC matrix elements are derived by dividing out phase space factors. Assuming lepton universality, these empirical matrix elements are used to calculate the ($e^-,\nue$) and  ($\bn_e,e^+$) amplitudes required for the LDCE calculations. The entrance channel ($e^-,\nue$) amplitudes and the related total cross sections for CC reactions on a $^{208}$Pb target are shown in Fig.~\ref{fig:enu}. As found in \cite{Formaggio:2012cpf} for $\nu_\mu$-nucleus reactions, the prominent sequence of QE, RE and DI components is visible,  strikingly resembling a (dissipated) three-level gross structure. 

The QE, RE, and DI matrix elements reflect the respective integrated cumulative nuclear/hadronic spectroscopic strengths in the out-channels as probed by the impinging lepton at given energy. Thus, the number of effectively available CC exit channels is estimated separately for $f\in \{QE,RE,DI\}$ from the respective partial mean matrix elements, similar to the decomposition in Fig.\ref{fig:enu}:  
\be \label{eq:NoS}
N^{(\bn_e,e^+)}_f(E)=  \frac{\lan |\mathcal{M}^{(\bn_e, e^+)}_{CX_f}(E)|^2\ran}{|\widetilde{G}_F|^2}\equiv \int_{0}^{E}d\omega S_{X_f}(\omega),
\ee
related to spectral distributions $S_{X_f}(E)$  in the final hadronic states $X_f$ where $E=M_{X_f}-M_B$.
Moreover, we have to account for the intrinsic two-body nature of the DCC interactions which is not accounted for properly by using the CC matrix elements originating from one-body operators only. Simple counting of nucleons, inspired by QE processes, leads to a factor $\frac{1}{2}A(A-1)$, which probably represents a lower limit for two-body enhancement in view of the additional degrees of freedom encountered in RE and DI processes.  

The phenomenological (on-shell) CC amplitudes are used as input for schematic LDCE calculations. The no-recoil limit is used which amounts to assume $M_C\gg m_i$.
Numerically, at beam energy $T_{lab} \simeq 1$~GeV the second order LDCE amplitudes are strongly dominated by the pole terms of the propagators, indicating that the reactions proceed effectively under on-shell conditions. At the energies of interest, off-shell effects are contributing by one percent or less. 
Hence, we have at hand a phenomenological approach which is expected to fully account for the experimentally confirmed total CC strengths in the QE, RE, and DI regions, albeit being limited to total cross sections. 

The limitations of the approach would be unduly exhausted if attempting a full-scale LDCE calculation. Rather, as \textit{pars pro toto} we investigate energy and mass dependence of the energy-momentum LR-mixing components which are expected to dominate LDCE reactions at multi-GeV energies. The calculations take into account the full spectrum of intermediate states $C(Z_\alpha,B_\alpha)$ and the complete complex-valued propagator in no-recoil approximation. The spectroscopic strength in the exit channel is treated by Eq.\eqref{eq:NoS}, including also two-body enhancement.

\begin{figure}
    \centering
    \includegraphics[width=1.0\linewidth]{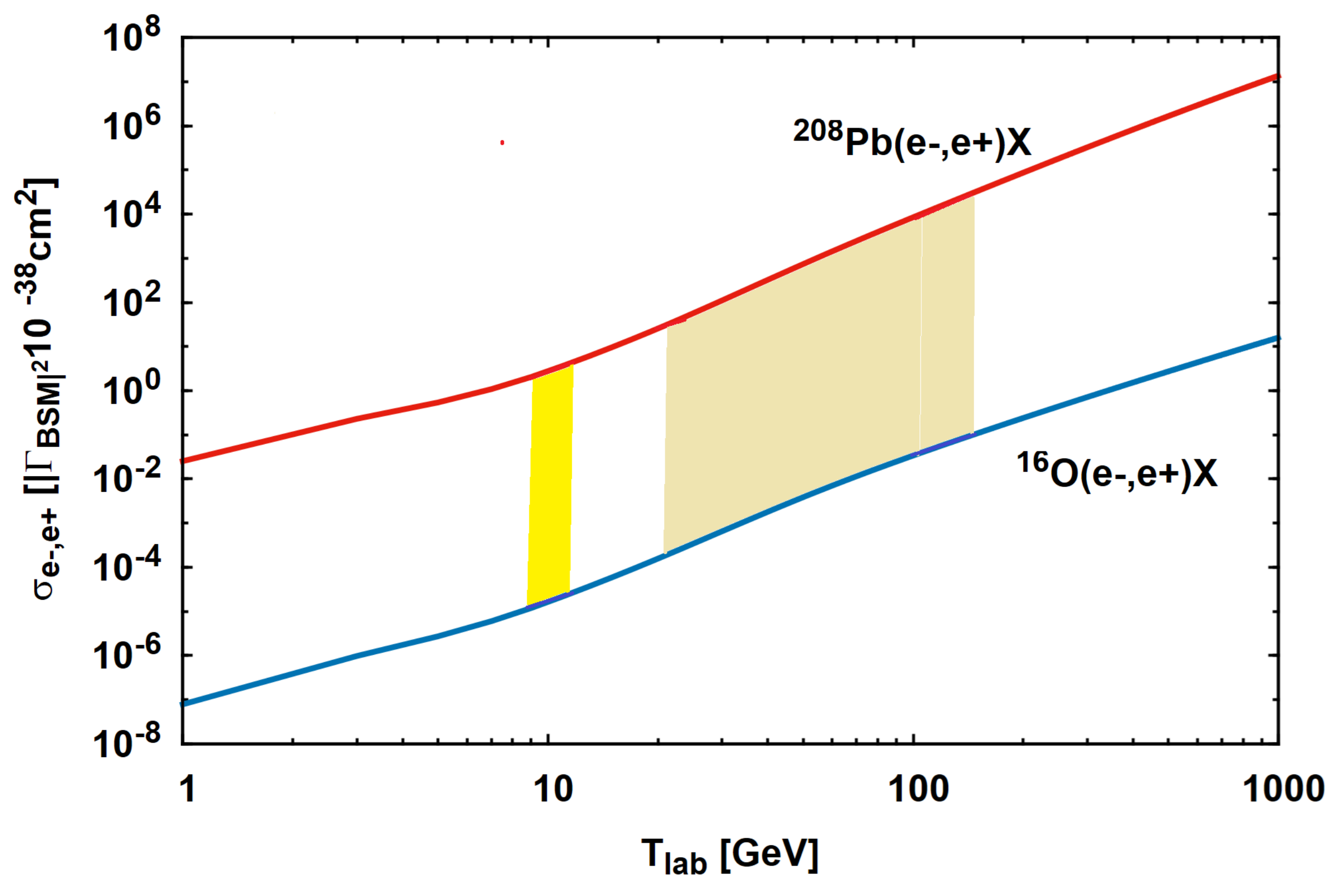}
    \caption{Schematic $(e^-,e^+)$ cross-sections on a $^{16}$O and a $^{208}$Pb target as a functions of the incident beam energy, representative for the two L/R mixing components (see text). JLab (yellow) and EIC (light yellow) energy regions are marked. An uncertainty of at least 20\% is expected already from the input CC matrix elements plus hardly to specify systematic contributions.}
    \label{fig:xsec_vs_energy}
\end{figure}

LDCE calculations have been performed for reactions on nuclei from $^ {16}$O to $^{208}$Pb.
In Fig.~\ref{fig:xsec_vs_energy}, the inclusive total cross section is shown for the LDCE reactions $^{16}$O($e^-,e^+$)X  and $^{208}$Pb($e^-,e^+$)X as function of the beam energy in units of the unknown strengths $|\Gamma_{BSM}|^2\in\{\eta,\kappa,\lambda\}\otimes\{\eta,\kappa,\lambda\} $. 
The LDCE cross section on $^{208}$Pb at $T_{lab}\approx$~10~GeV is already sizable with a strength of close to
$10^{-37}\times|\Gamma_{BSM}|^2  cm^2$ (or equivalently $100\times|\Gamma_{BSM}|^2$ fb). That value encourages first feasibility
experiments, provided that sufficient luminosity is available.
The total cross sections depend on beam energy by a power law $\sim T^3_{lab}$.

\begin{figure}
    \centering
    \includegraphics[width=0.95\linewidth]{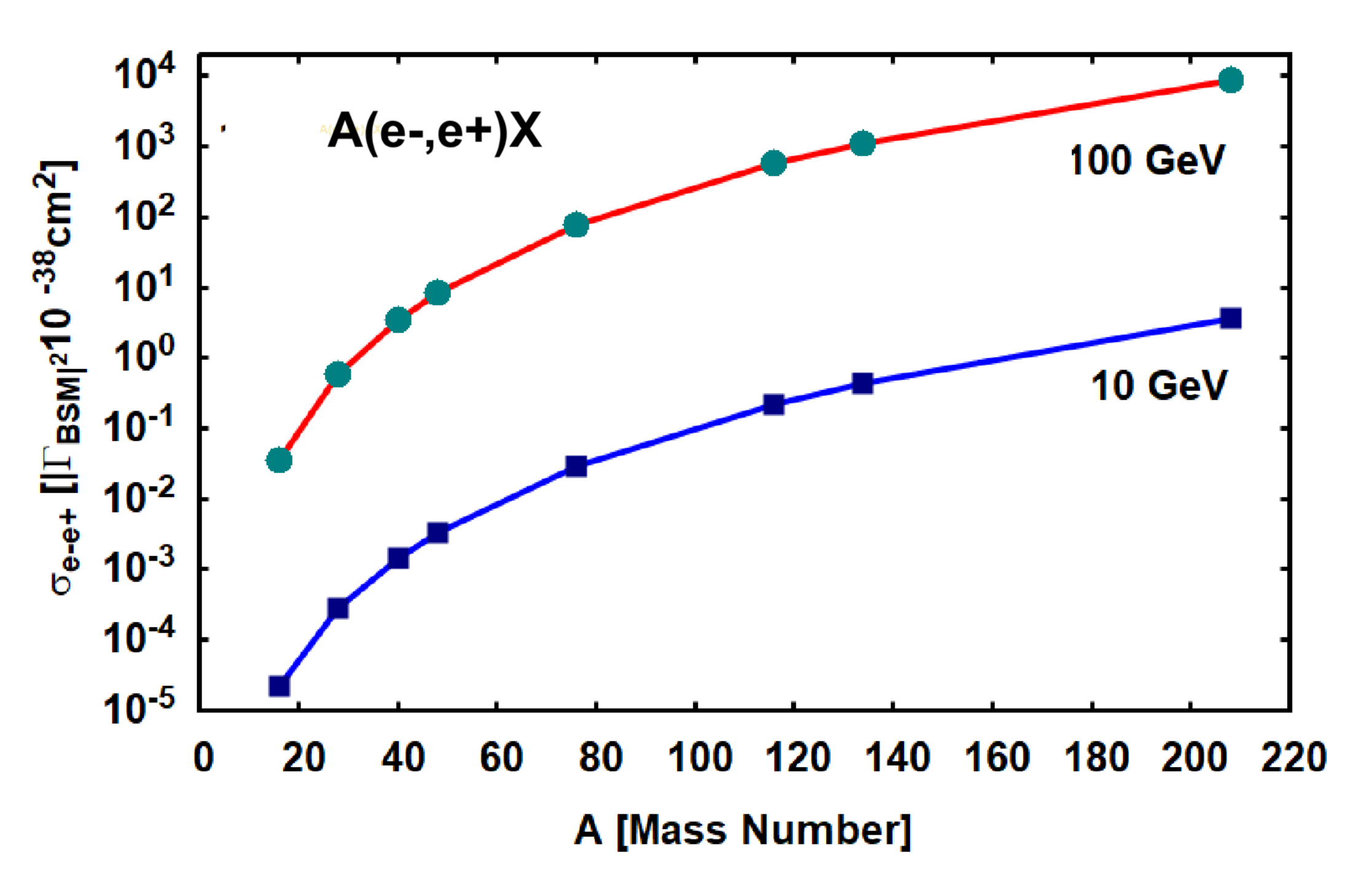}
    \caption{Mass dependence of the total LDCE cross section predicted by the second-order phenomenological approach. Results at $T_{lab}=10$~GeV and $T_{lab}=100$~GeV for a representative selection of nuclei of potential experimental interest are indicated by symbols. The lines are drawn to guide the eye.}
    \label{fig:xsec_vs_mass}
\end{figure}

In Fig.~\ref{fig:xsec_vs_mass}, the mass dependence is illustrated for two beam energies, the one relevant for the existing Jefferson Laboratory (JLab) facility, the other of potential interest for studies at the Electron Ion Collider (EIC). The yields increase rapidly with mass number. The clear preference of heavy target nuclei in the mass range above $A > 100$ is obvious.

\paragraph{\textbf{Experimental Issues and Outlook:}}
The feasibility of LDCE reactions at existing facilities as JLab depends of course on a multitude of parameters which are not accessible directly in our preliminary calculations which reflect only the prompt positron channel. However, a few general aspects are worth mentioning.   
In LDCE reactions at about 10 GeV incident energy, the prompt $e^+$ ejectiles are expected to be emitted preferentially at forward angles $(5^\circ - 15^\circ)$ with energies close to the beam energy. This resembles the situation encountered in $\nu_\mu + X(Z,N)$ and $\nu_e + X(Z,N)$ CC reactions at similar average neutrino beam energies \cite{Formaggio:2012cpf}. Under such kinematical condition, the background from energetic  multiple $e^+$ electromagnetic interactions is strongly attenuated, but will be still much larger than the expected $e^+$ signal rate from LDCE. Therefore, detection of energetic $e^+$ at forward angles by itself will not be a sufficient signature for a LDCE reaction. Rather, it is crucial to obtain additional information from the detection of reaction products allowing to encircle the final nuclear configuration. Hence, as a minimal requirement  other reaction products should be recorded in coincidence with the $e^+$. The proper type of fragments will depend on the energy region, thus changing from QE to RE and DI energies. Detection of the knocked-out particles in coincidence with the positron can be used to define unambiguous signatures of the LDCE events.

Momentum conservation leads to a specific momentum transfer $Q^2$ with a well-defined emission direction for the center-of-mass of the set of emitted particles. The $\nu_\mu + X(Z,N)$ and $\nu_e + X(Z,N)$ CC differential cross sections data indicate $Q^2 \approx 1- 2\,GeV^2/c^2$ and angles $\theta_h (\nu_{e,\mu} +X(Z,N)) \approx 30^\circ$ for the detected hadrons, mostly pions \cite{lozano:2025Minerva,Acero:2023Nova,Lu:2021Minerva}. Since in the intermediate energy region up to about 10 GeV the DI, RE and  QE mechanisms compete, it is important to explore as many outgoing channels as possible in exclusive  measurements, requiring multiple coincidence measurements. This, in turn, calls for rather high detection efficiency for different hadron species, unambiguous particle identification and good angular resolution to enable the reconstruction of the hadron showers.

Although the electron-positron conversion process may be generalized to any kind of incoming and outgoing charged leptons  $(l^\pm,l'^\mp )\in\{ e^\pm ,\mu^\pm ,\tau^\pm \}$, electron beams clearly offer the best solution in terms of availability, high luminosities, variable energies, high momentum resolution and low emittance. Thus, electron accelerators facilities allow to explore systematically large sectors of the energy-momentum space, possibly also with polarized beams or targets which even would allow to study the full operator structure of the LDCE amplitude, Eq.\eqref{eq:MeeSquared}. Among the existing facilities, the 12 GeV electron beams with up to $i=100\,\mu A$ current and targets with areal density $A \approx 1\,g/cm^2$ available at JLab seem to be most promising by reaching luminosities which, in principle, should allow to observe LDCE reactions with sensitivities down to $|\Gamma_{BSM}|^2 \approx 10^{-6}$. Even better perspectives can be expected for experiments in
collider mode with crossing beams of electrons and heavy ions, as planned at the EIC. The EIC is designed to reach beam-energies up to 140 GeV in the center-of-mass frame, albeit with a smaller luminosity of about $10^{31-32}  cm^{-2} s^{-1}$ for a $^{208}Pb$ target, gaining approximately 4 orders of magnitude increase in the LDCE cross section.
The beam current is not the only factor to be considered. A realistic assessment must also account for possible limitations arising from the target, the detection systems, the data acquisition chain, and related components. 

An important task of future research is to consider recoil effects which were neglected in the exploratory numerical studies. Recoil effects will become important if HNL should exist with masses comparable to nuclear masses,  $m_{HNL}\gtrsim M_C$. Varying the target masses may give already important hints because HNL will produce observable effects by changing the eigenspectrum of the intermediate channels. Moreover, HNL will not only affecting LDCE cross sections but probably indirectly also the environment in which the target nucleus is embedded.       

In conclusion we propose a new approach to LNV research utilizing ($e^-,e^+$) LDCE reactions on nuclear targets at accelerator facilities. Based on the LRSM, the reactions are described as second order CC processes, 
the cross sections are dominated by energy-momentum dependent LR-mixing contributions. At the beam energies of interest the LNL Majorana mass terms are strongly suppressed. 
The LDCE approach is minimalistic in the sense that besides the LRSM no further assumptions on BSM mechanisms have to be invented.
LDCE reactions are determined by effective dim-5 operators, similar but not identical to the DBD operators. Clearly, operators of higher complexity might contribute, too. Additional ($e^-,e^+$) transition strengths can be expected from higher-dimensional operators which will compete and interfere with the LDCE amplitudes. Under reaction-theoretical aspects those operators lead to first order RME for which a different energy and momentum dependence is expected. 
While only a minor impact of high-dim operators was found in recent DBD studies \cite{Ejiri:2025oro}, considerable enhancements are expected, but not proven, in the LLjj high energy sector \cite{Barry:2013xxa,Fridell:2023rtr}. Until now, nothing is known on high-dimensional LNV operators in the multi-GeV region, a gap that can be filled naturally by accelerator LNV studies.

First pioneering explorations of LDCE reactions could be performed timely at existing facilities, such as JLab, with multi-GeV electron beams and the proper detection technologies for inclusive measurement of reaction products. Even the absence of LDCE events would produce significant advances in LNV research by establishing upper LNV-limits in a hitherto unexplored energy region and stimulating dedicated second-generation experiments. 
Conversely, the observation of only a few unequivocal events would be already a major scientific breakthrough.

\begin{acknowledgments}
H.L. acknowledges financial support in part by DFG, grants Le439/16 and Le439/17.
\end{acknowledgments}

%\bibliographystyle{apsrev4-1}
%\bibliography{LDCE_2026}

\begin{thebibliography}{31}%
\makeatletter
\providecommand \@ifxundefined [1]{%
 \@ifx{#1\undefined}
}%
\providecommand \@ifnum [1]{%
 \ifnum #1\expandafter \@firstoftwo
 \else \expandafter \@secondoftwo
 \fi
}%
\providecommand \@ifx [1]{%
 \ifx #1\expandafter \@firstoftwo
 \else \expandafter \@secondoftwo
 \fi
}%
\providecommand \natexlab [1]{#1}%
\providecommand \enquote  [1]{``#1''}%
\providecommand \bibnamefont  [1]{#1}%
\providecommand \bibfnamefont [1]{#1}%
\providecommand \citenamefont [1]{#1}%
\providecommand \href@noop [0]{\@secondoftwo}%
\providecommand \href [0]{\begingroup \@sanitize@url \@href}%
\providecommand \@href[1]{\@@startlink{#1}\@@href}%
\providecommand \@@href[1]{\endgroup#1\@@endlink}%
\providecommand \@sanitize@url [0]{\catcode `\\12\catcode `\$12\catcode
  `\&12\catcode `\#12\catcode `\^12\catcode `\_12\catcode `\%12\relax}%
\providecommand \@@startlink[1]{}%
\providecommand \@@endlink[0]{}%
\providecommand \url  [0]{\begingroup\@sanitize@url \@url }%
\providecommand \@url [1]{\endgroup\@href {#1}{\urlprefix }}%
\providecommand \urlprefix  [0]{URL }%
\providecommand \Eprint [0]{\href }%
\providecommand \doibase [0]{http://dx.doi.org/}%
\providecommand \selectlanguage [0]{\@gobble}%
\providecommand \bibinfo  [0]{\@secondoftwo}%
\providecommand \bibfield  [0]{\@secondoftwo}%
\providecommand \translation [1]{[#1]}%
\providecommand \BibitemOpen [0]{}%
\providecommand \bibitemStop [0]{}%
\providecommand \bibitemNoStop [0]{.\EOS\space}%
\providecommand \EOS [0]{\spacefactor3000\relax}%
\providecommand \BibitemShut  [1]{\csname bibitem#1\endcsname}%
\let\auto@bib@innerbib\@empty
%</preamble>
\bibitem [{\citenamefont {Ahmad}\ \emph {et~al.}(1988)\citenamefont {Ahmad},
  \citenamefont {Azuelos}, \citenamefont {Blecher}, \citenamefont {Bryman},
  \citenamefont {Burnham}, \citenamefont {Clifford}, \citenamefont {Depommier},
  \citenamefont {Dixit}, \citenamefont {Gotow}, \citenamefont {Hargrove},
  \citenamefont {Hasinoff}, \citenamefont {Leitch}, \citenamefont {Macdonald},
  \citenamefont {Mes}, \citenamefont {Navon}, \citenamefont {Numao},
  \citenamefont {Poutissou}, \citenamefont {Poutissou}, \citenamefont
  {Schlatter}, \citenamefont {Spuller},\ and\ \citenamefont
  {Summhammer}}]{PhysRevD.38.2102}%
  \BibitemOpen
  \bibfield  {author} {\bibinfo {author} {\bibfnamefont {S.}~\bibnamefont
  {Ahmad}}, \bibinfo {author} {\bibfnamefont {G.}~\bibnamefont {Azuelos}},
  \bibinfo {author} {\bibfnamefont {M.}~\bibnamefont {Blecher}}, \bibinfo
  {author} {\bibfnamefont {D.~A.}\ \bibnamefont {Bryman}}, \bibinfo {author}
  {\bibfnamefont {R.~A.}\ \bibnamefont {Burnham}}, \bibinfo {author}
  {\bibfnamefont {E.~T.~H.}\ \bibnamefont {Clifford}}, \bibinfo {author}
  {\bibfnamefont {P.}~\bibnamefont {Depommier}}, \bibinfo {author}
  {\bibfnamefont {M.~S.}\ \bibnamefont {Dixit}}, \bibinfo {author}
  {\bibfnamefont {K.}~\bibnamefont {Gotow}}, \bibinfo {author} {\bibfnamefont
  {C.~K.}\ \bibnamefont {Hargrove}}, \bibinfo {author} {\bibfnamefont
  {M.}~\bibnamefont {Hasinoff}}, \bibinfo {author} {\bibfnamefont
  {M.}~\bibnamefont {Leitch}}, \bibinfo {author} {\bibfnamefont {J.~A.}\
  \bibnamefont {Macdonald}}, \bibinfo {author} {\bibfnamefont {H.}~\bibnamefont
  {Mes}}, \bibinfo {author} {\bibfnamefont {I.}~\bibnamefont {Navon}}, \bibinfo
  {author} {\bibfnamefont {T.}~\bibnamefont {Numao}}, \bibinfo {author}
  {\bibfnamefont {J.-M.}\ \bibnamefont {Poutissou}}, \bibinfo {author}
  {\bibfnamefont {R.}~\bibnamefont {Poutissou}}, \bibinfo {author}
  {\bibfnamefont {P.}~\bibnamefont {Schlatter}}, \bibinfo {author}
  {\bibfnamefont {J.}~\bibnamefont {Spuller}}, \ and\ \bibinfo {author}
  {\bibfnamefont {J.}~\bibnamefont {Summhammer}},\ }\href {\doibase
  10.1103/PhysRevD.38.2102} {\bibfield  {journal} {\bibinfo  {journal} {Phys.
  Rev. D}\ }\textbf {\bibinfo {volume} {38}},\ \bibinfo {pages} {2102}
  (\bibinfo {year} {1988})}\BibitemShut {NoStop}%
\bibitem [{\citenamefont {Vergados}(1982)}]{Vergados:1981bm}%
  \BibitemOpen
  \bibfield  {author} {\bibinfo {author} {\bibfnamefont {J.~D.}\ \bibnamefont
  {Vergados}},\ }\href {\doibase 10.1103/PhysRevD.25.914} {\bibfield  {journal}
  {\bibinfo  {journal} {Phys. Rev. D}\ }\textbf {\bibinfo {volume} {25}},\
  \bibinfo {pages} {914} (\bibinfo {year} {1982})}\BibitemShut {NoStop}%
\bibitem [{\citenamefont {Vergados}(2002)}]{Vergados:2002pv}%
  \BibitemOpen
  \bibfield  {author} {\bibinfo {author} {\bibfnamefont {J.~D.}\ \bibnamefont
  {Vergados}},\ }\href {\doibase 10.1016/S0370-1573(01)00068-0} {\bibfield
  {journal} {\bibinfo  {journal} {Phys. Rept.}\ }\textbf {\bibinfo {volume}
  {361}},\ \bibinfo {pages} {1} (\bibinfo {year} {2002})},\ \Eprint
  {http://arxiv.org/abs/hep-ph/0209347} {arXiv:hep-ph/0209347} \BibitemShut
  {NoStop}%
\bibitem [{\citenamefont {Divari}\ \emph {et~al.}(2002)\citenamefont {Divari},
  \citenamefont {Vergados}, \citenamefont {Kosmas},\ and\ \citenamefont
  {Skouras}}]{Divari:2002sq}%
  \BibitemOpen
  \bibfield  {author} {\bibinfo {author} {\bibfnamefont {P.~C.}\ \bibnamefont
  {Divari}}, \bibinfo {author} {\bibfnamefont {J.~D.}\ \bibnamefont
  {Vergados}}, \bibinfo {author} {\bibfnamefont {T.~S.}\ \bibnamefont
  {Kosmas}}, \ and\ \bibinfo {author} {\bibfnamefont {L.~D.}\ \bibnamefont
  {Skouras}},\ }\href {\doibase 10.1016/S0375-9474(01)01533-0} {\bibfield
  {journal} {\bibinfo  {journal} {Nucl. Phys. A}\ }\textbf {\bibinfo {volume}
  {703}},\ \bibinfo {pages} {409} (\bibinfo {year} {2002})},\ \Eprint
  {http://arxiv.org/abs/nucl-th/0203066} {arXiv:nucl-th/0203066} \BibitemShut
  {NoStop}%
\bibitem [{\citenamefont {Missimer}\ \emph {et~al.}(1994)\citenamefont
  {Missimer}, \citenamefont {Mohapatra},\ and\ \citenamefont
  {Mukhopadhyay}}]{Missimer:1994xd}%
  \BibitemOpen
  \bibfield  {author} {\bibinfo {author} {\bibfnamefont {J.~H.}\ \bibnamefont
  {Missimer}}, \bibinfo {author} {\bibfnamefont {R.~N.}\ \bibnamefont
  {Mohapatra}}, \ and\ \bibinfo {author} {\bibfnamefont {N.~C.}\ \bibnamefont
  {Mukhopadhyay}},\ }\href {\doibase 10.1103/PhysRevD.50.2067} {\bibfield
  {journal} {\bibinfo  {journal} {Phys. Rev. D}\ }\textbf {\bibinfo {volume}
  {50}},\ \bibinfo {pages} {2067} (\bibinfo {year} {1994})}\BibitemShut
  {NoStop}%
\bibitem [{\citenamefont {Badertscher}\ \emph {et~al.}(1991)\citenamefont
  {Badertscher} \emph {et~al.}}]{SINDRUMII:1991lmr}%
  \BibitemOpen
  \bibfield  {author} {\bibinfo {author} {\bibfnamefont {A.}~\bibnamefont
  {Badertscher}} \emph {et~al.} (\bibinfo {collaboration} {SINDRUM II}),\
  }\href {\doibase 10.1088/0954-3899/17/S/005} {\bibfield  {journal} {\bibinfo
  {journal} {J. Phys. G}\ }\textbf {\bibinfo {volume} {17}},\ \bibinfo {pages}
  {S47} (\bibinfo {year} {1991})}\BibitemShut {NoStop}%
\bibitem [{\citenamefont {Bertl}\ \emph {et~al.}(2006)\citenamefont {Bertl}
  \emph {et~al.}}]{SINDRUMII:2006dvw}%
  \BibitemOpen
  \bibfield  {author} {\bibinfo {author} {\bibfnamefont {W.~H.}\ \bibnamefont
  {Bertl}} \emph {et~al.} (\bibinfo {collaboration} {SINDRUM II}),\ }\href
  {\doibase 10.1140/epjc/s2006-02582-x} {\bibfield  {journal} {\bibinfo
  {journal} {Eur. Phys. J. C}\ }\textbf {\bibinfo {volume} {47}},\ \bibinfo
  {pages} {337} (\bibinfo {year} {2006})}\BibitemShut {NoStop}%
\bibitem [{\citenamefont {van~der Schaaf}(2021)}]{vanderSchaaf:2021hnd}%
  \BibitemOpen
  \bibfield  {author} {\bibinfo {author} {\bibfnamefont {A.}~\bibnamefont
  {van~der Schaaf}},\ }\href {\doibase 10.21468/SciPostPhysProc.5.008}
  {\bibfield  {journal} {\bibinfo  {journal} {SciPost Phys. Proc.}\ }\textbf
  {\bibinfo {volume} {5}},\ \bibinfo {pages} {008} (\bibinfo {year}
  {2021})}\BibitemShut {NoStop}%
\bibitem [{\citenamefont {Babu}\ \emph {et~al.}(2024)\citenamefont {Babu},
  \citenamefont {Barman}, \citenamefont {Gon\c{c}alves},\ and\ \citenamefont
  {Ismail}}]{Babu:2022ycv}%
  \BibitemOpen
  \bibfield  {author} {\bibinfo {author} {\bibfnamefont {K.~S.}\ \bibnamefont
  {Babu}}, \bibinfo {author} {\bibfnamefont {R.~K.}\ \bibnamefont {Barman}},
  \bibinfo {author} {\bibfnamefont {D.}~\bibnamefont {Gon\c{c}alves}}, \ and\
  \bibinfo {author} {\bibfnamefont {A.}~\bibnamefont {Ismail}},\ }\href
  {\doibase 10.1007/JHEP06(2024)132} {\bibfield  {journal} {\bibinfo  {journal}
  {JHEP}\ }\textbf {\bibinfo {volume} {06}},\ \bibinfo {pages} {132} (\bibinfo
  {year} {2024})},\ \Eprint {http://arxiv.org/abs/2212.08025} {arXiv:2212.08025
  [hep-ph]} \BibitemShut {NoStop}%
\bibitem [{\citenamefont {Mikulenko}\ and\ \citenamefont
  {Marinichenko}(2024)}]{Mikulenko:2023ezx}%
  \BibitemOpen
  \bibfield  {author} {\bibinfo {author} {\bibfnamefont {O.}~\bibnamefont
  {Mikulenko}}\ and\ \bibinfo {author} {\bibfnamefont {M.}~\bibnamefont
  {Marinichenko}},\ }\href {\doibase 10.1007/JHEP01(2024)032} {\bibfield
  {journal} {\bibinfo  {journal} {JHEP}\ }\textbf {\bibinfo {volume} {01}},\
  \bibinfo {pages} {032} (\bibinfo {year} {2024})},\ \Eprint
  {http://arxiv.org/abs/2309.16837} {arXiv:2309.16837 [hep-ph]} \BibitemShut
  {NoStop}%
\bibitem [{\citenamefont {Yang}\ \emph {et~al.}(2025)\citenamefont {Yang},
  \citenamefont {Shao}, \citenamefont {Han}, \citenamefont {Huang},
  \citenamefont {Jin},\ and\ \citenamefont {Li}}]{Yang:2025jxc}%
  \BibitemOpen
  \bibfield  {author} {\bibinfo {author} {\bibfnamefont {F.~X.}\ \bibnamefont
  {Yang}}, \bibinfo {author} {\bibfnamefont {F.~L.}\ \bibnamefont {Shao}},
  \bibinfo {author} {\bibfnamefont {Z.~L.}\ \bibnamefont {Han}}, \bibinfo
  {author} {\bibfnamefont {F.}~\bibnamefont {Huang}}, \bibinfo {author}
  {\bibfnamefont {Y.}~\bibnamefont {Jin}}, \ and\ \bibinfo {author}
  {\bibfnamefont {H.}~\bibnamefont {Li}},\ }\href {\doibase
  10.1007/JHEP09(2025)200} {\bibfield  {journal} {\bibinfo  {journal} {JHEP}\
  }\textbf {\bibinfo {volume} {09}},\ \bibinfo {pages} {200} (\bibinfo {year}
  {2025})},\ \Eprint {http://arxiv.org/abs/2505.07331} {arXiv:2505.07331
  [hep-ph]} \BibitemShut {NoStop}%
\bibitem [{\citenamefont {Liao}\ and\ \citenamefont {Ma}(2016)}]{Liao:2016hru}%
  \BibitemOpen
  \bibfield  {author} {\bibinfo {author} {\bibfnamefont {Y.}~\bibnamefont
  {Liao}}\ and\ \bibinfo {author} {\bibfnamefont {X.~D.}\ \bibnamefont {Ma}},\
  }\href {\doibase 10.1007/JHEP11(2016)043} {\bibfield  {journal} {\bibinfo
  {journal} {JHEP}\ }\textbf {\bibinfo {volume} {11}},\ \bibinfo {pages} {043}
  (\bibinfo {year} {2016})},\ \Eprint {http://arxiv.org/abs/1607.07309}
  {arXiv:1607.07309 [hep-ph]} \BibitemShut {NoStop}%
\bibitem [{\citenamefont {Barry}\ and\ \citenamefont
  {Rodejohann}(2013)}]{Barry:2013xxa}%
  \BibitemOpen
  \bibfield  {author} {\bibinfo {author} {\bibfnamefont {J.}~\bibnamefont
  {Barry}}\ and\ \bibinfo {author} {\bibfnamefont {W.}~\bibnamefont
  {Rodejohann}},\ }\href {\doibase 10.1007/JHEP09(2013)153} {\bibfield
  {journal} {\bibinfo  {journal} {JHEP}\ }\textbf {\bibinfo {volume} {09}},\
  \bibinfo {pages} {153} (\bibinfo {year} {2013})},\ \Eprint
  {http://arxiv.org/abs/1303.6324} {arXiv:1303.6324 [hep-ph]} \BibitemShut
  {NoStop}%
\bibitem [{\citenamefont {Fridell}\ \emph {et~al.}(2024)\citenamefont
  {Fridell}, \citenamefont {Gr\'af}, \citenamefont {Harz},\ and\ \citenamefont
  {Hati}}]{Fridell:2023rtr}%
  \BibitemOpen
  \bibfield  {author} {\bibinfo {author} {\bibfnamefont {K.}~\bibnamefont
  {Fridell}}, \bibinfo {author} {\bibfnamefont {L.}~\bibnamefont {Gr\'af}},
  \bibinfo {author} {\bibfnamefont {J.}~\bibnamefont {Harz}}, \ and\ \bibinfo
  {author} {\bibfnamefont {C.}~\bibnamefont {Hati}},\ }\href {\doibase
  10.1007/JHEP05(2024)154} {\bibfield  {journal} {\bibinfo  {journal} {JHEP}\
  }\textbf {\bibinfo {volume} {05}},\ \bibinfo {pages} {154} (\bibinfo {year}
  {2024})},\ \Eprint {http://arxiv.org/abs/2306.08709} {arXiv:2306.08709
  [hep-ph]} \BibitemShut {NoStop}%
\bibitem [{\citenamefont {Antusch}\ \emph {et~al.}(2023)\citenamefont
  {Antusch}, \citenamefont {Hajer},\ and\ \citenamefont
  {Oliveira}}]{Antusch:2023jsa}%
  \BibitemOpen
  \bibfield  {author} {\bibinfo {author} {\bibfnamefont {S.}~\bibnamefont
  {Antusch}}, \bibinfo {author} {\bibfnamefont {J.}~\bibnamefont {Hajer}}, \
  and\ \bibinfo {author} {\bibfnamefont {B.~M.~S.}\ \bibnamefont {Oliveira}},\
  }\href {\doibase 10.1007/JHEP10(2023)129} {\bibfield  {journal} {\bibinfo
  {journal} {JHEP}\ }\textbf {\bibinfo {volume} {10}},\ \bibinfo {pages} {129}
  (\bibinfo {year} {2023})},\ \Eprint {http://arxiv.org/abs/2308.07297}
  {arXiv:2308.07297 [hep-ph]} \BibitemShut {NoStop}%
\bibitem [{\citenamefont {Schechter}\ and\ \citenamefont
  {Valle}(1982)}]{Schechter:1982dbd}%
  \BibitemOpen
  \bibfield  {author} {\bibinfo {author} {\bibfnamefont {J.}~\bibnamefont
  {Schechter}}\ and\ \bibinfo {author} {\bibfnamefont {J.~W.~F.}\ \bibnamefont
  {Valle}},\ }\href {\doibase 10.1103/PhysRevD.25.2951} {\bibfield  {journal}
  {\bibinfo  {journal} {Phys. Rev. D}\ }\textbf {\bibinfo {volume} {25}},\
  \bibinfo {pages} {2951} (\bibinfo {year} {1982})}\BibitemShut {NoStop}%
\bibitem [{\citenamefont {Altarelli}\ and\ \citenamefont
  {Winter}(2003)}]{Altarelli:2003ph}%
  \BibitemOpen
  \bibinfo {editor} {\bibfnamefont {G.}~\bibnamefont {Altarelli}}\ and\
  \bibinfo {editor} {\bibfnamefont {K.}~\bibnamefont {Winter}},\ eds.,\ \href
  {\doibase 10.1007/b13585} {\emph {\bibinfo {title} {{Neutrino mass}}}},\
  Vol.\ \bibinfo {volume} {190}\ (\bibinfo  {publisher} {Springer Tracts in
  Mod. Phys.},\ \bibinfo {year} {2003})\BibitemShut {NoStop}%
\bibitem [{\citenamefont {Buchmuller}\ \emph {et~al.}(2005)\citenamefont
  {Buchmuller}, \citenamefont {Di~Bari},\ and\ \citenamefont
  {Plumacher}}]{Buchmuller:2004nz}%
  \BibitemOpen
  \bibfield  {author} {\bibinfo {author} {\bibfnamefont {W.}~\bibnamefont
  {Buchmuller}}, \bibinfo {author} {\bibfnamefont {P.}~\bibnamefont {Di~Bari}},
  \ and\ \bibinfo {author} {\bibfnamefont {M.}~\bibnamefont {Plumacher}},\
  }\href {\doibase 10.1016/j.aop.2004.02.003} {\bibfield  {journal} {\bibinfo
  {journal} {Annals Phys.}\ }\textbf {\bibinfo {volume} {315}},\ \bibinfo
  {pages} {305} (\bibinfo {year} {2005})},\ \Eprint
  {http://arxiv.org/abs/hep-ph/0401240} {arXiv:hep-ph/0401240} \BibitemShut
  {NoStop}%
\bibitem [{\citenamefont {Kayser}(2008)}]{Kayser:2008rd}%
  \BibitemOpen
  \bibfield  {author} {\bibinfo {author} {\bibfnamefont {B.}~\bibnamefont
  {Kayser}},\ }\href {\doibase 10.48550/arXiv.0804.1497} {\  (\bibinfo {year}
  {2008}),\ 10.48550/arXiv.0804.1497},\ \Eprint
  {http://arxiv.org/abs/0804.1497} {arXiv:0804.1497 [hep-ph]} \BibitemShut
  {NoStop}%
\bibitem [{\citenamefont {Doi}\ \emph {et~al.}(1983)\citenamefont {Doi},
  \citenamefont {Kotani}, \citenamefont {Nishiura},\ and\ \citenamefont
  {Takasugi}}]{Doi:1982dn}%
  \BibitemOpen
  \bibfield  {author} {\bibinfo {author} {\bibfnamefont {M.}~\bibnamefont
  {Doi}}, \bibinfo {author} {\bibfnamefont {T.}~\bibnamefont {Kotani}},
  \bibinfo {author} {\bibfnamefont {H.}~\bibnamefont {Nishiura}}, \ and\
  \bibinfo {author} {\bibfnamefont {E.}~\bibnamefont {Takasugi}},\ }\href
  {\doibase 10.1143/PTP.69.602} {\bibfield  {journal} {\bibinfo  {journal}
  {Prog. Theor. Phys.}\ }\textbf {\bibinfo {volume} {69}},\ \bibinfo {pages}
  {602} (\bibinfo {year} {1983})}\BibitemShut {NoStop}%
\bibitem [{\citenamefont {Tomoda}(1991)}]{Tomoda:1990rs}%
  \BibitemOpen
  \bibfield  {author} {\bibinfo {author} {\bibfnamefont {T.}~\bibnamefont
  {Tomoda}},\ }\href {\doibase doi:10.1088/0034-4885/54/1/002} {\bibfield
  {journal} {\bibinfo  {journal} {Rept. Prog. Phys.}\ }\textbf {\bibinfo
  {volume} {54}},\ \bibinfo {pages} {53} (\bibinfo {year} {1991})}\BibitemShut
  {NoStop}%
\bibitem [{\citenamefont {Vergados}\ \emph {et~al.}(2012)\citenamefont
  {Vergados}, \citenamefont {Ejiri},\ and\ \citenamefont
  {Simkovic}}]{Vergados:2012xy}%
  \BibitemOpen
  \bibfield  {author} {\bibinfo {author} {\bibfnamefont {J.~D.}\ \bibnamefont
  {Vergados}}, \bibinfo {author} {\bibfnamefont {H.}~\bibnamefont {Ejiri}}, \
  and\ \bibinfo {author} {\bibfnamefont {F.}~\bibnamefont {Simkovic}},\ }\href
  {\doibase 10.1088/0034-4885/75/10/106301} {\bibfield  {journal} {\bibinfo
  {journal} {Rept. Prog. Phys.}\ }\textbf {\bibinfo {volume} {75}},\ \bibinfo
  {pages} {106301} (\bibinfo {year} {2012})},\ \Eprint
  {http://arxiv.org/abs/1205.0649} {arXiv:1205.0649 [hep-ph]} \BibitemShut
  {NoStop}%
%%CITATION = ARXIV:1205.0649;%%
\bibitem [{\citenamefont {Pontecorvo}(1957)}]{Pontecorvo:1957qd}%
  \BibitemOpen
  \bibfield  {author} {\bibinfo {author} {\bibfnamefont {B.}~\bibnamefont
  {Pontecorvo}},\ }\href@noop {} {\bibfield  {journal} {\bibinfo  {journal}
  {Zh. Eksp. Teor. Fiz.}\ }\textbf {\bibinfo {volume} {34}},\ \bibinfo {pages}
  {247} (\bibinfo {year} {1957})}\BibitemShut {NoStop}%
\bibitem [{\citenamefont {Maki}\ \emph {et~al.}(1960)\citenamefont {Maki},
  \citenamefont {Nakagawa}, \citenamefont {Ohnuki},\ and\ \citenamefont
  {Sakata}}]{Maki:1960ut}%
  \BibitemOpen
  \bibfield  {author} {\bibinfo {author} {\bibfnamefont {Z.}~\bibnamefont
  {Maki}}, \bibinfo {author} {\bibfnamefont {M.}~\bibnamefont {Nakagawa}},
  \bibinfo {author} {\bibfnamefont {Y.}~\bibnamefont {Ohnuki}}, \ and\ \bibinfo
  {author} {\bibfnamefont {S.}~\bibnamefont {Sakata}},\ }\href {\doibase
  10.1143/PTP.23.1174} {\bibfield  {journal} {\bibinfo  {journal} {Prog. Theor.
  Phys.}\ }\textbf {\bibinfo {volume} {23}},\ \bibinfo {pages} {1174} (\bibinfo
  {year} {1960})}\BibitemShut {NoStop}%
\bibitem [{\citenamefont {Formaggio}\ and\ \citenamefont
  {Zeller}(2012)}]{Formaggio:2012cpf}%
  \BibitemOpen
  \bibfield  {author} {\bibinfo {author} {\bibfnamefont {J.~A.}\ \bibnamefont
  {Formaggio}}\ and\ \bibinfo {author} {\bibfnamefont {G.~P.}\ \bibnamefont
  {Zeller}},\ }\href {\doibase 10.1103/RevModPhys.84.1307} {\bibfield
  {journal} {\bibinfo  {journal} {Rev. Mod. Phys.}\ }\textbf {\bibinfo {volume}
  {84}},\ \bibinfo {pages} {1307} (\bibinfo {year} {2012})},\ \Eprint
  {http://arxiv.org/abs/1305.7513} {arXiv:1305.7513 [hep-ex]} \BibitemShut
  {NoStop}%
\bibitem [{\citenamefont {Ejiri}\ \emph {et~al.}(2025)\citenamefont {Ejiri},
  \citenamefont {Fukuyama},\ and\ \citenamefont {Sato}}]{Ejiri:2025oro}%
  \BibitemOpen
  \bibfield  {author} {\bibinfo {author} {\bibfnamefont {H.}~\bibnamefont
  {Ejiri}}, \bibinfo {author} {\bibfnamefont {T.}~\bibnamefont {Fukuyama}}, \
  and\ \bibinfo {author} {\bibfnamefont {T.}~\bibnamefont {Sato}},\ }\href
  {\doibase 10.1103/f4cf-vnr9} {\bibfield  {journal} {\bibinfo  {journal}
  {Phys. Rev. C}\ }\textbf {\bibinfo {volume} {111}},\ \bibinfo {pages}
  {065501} (\bibinfo {year} {2025})},\ \Eprint
  {http://arxiv.org/abs/2501.03454} {arXiv:2501.03454 [hep-ph]} \BibitemShut
  {NoStop}%
\bibitem [{\citenamefont {{Sajjad Athar}}\ \emph {et~al.}(2023)\citenamefont
  {{Sajjad Athar}}, \citenamefont {Fatima},\ and\ \citenamefont
  {Singh}}]{SajjadAthar:2022pjt}%
  \BibitemOpen
  \bibfield  {author} {\bibinfo {author} {\bibfnamefont {M.}~\bibnamefont
  {{Sajjad Athar}}}, \bibinfo {author} {\bibfnamefont {A.}~\bibnamefont
  {Fatima}}, \ and\ \bibinfo {author} {\bibfnamefont {S.}~\bibnamefont
  {Singh}},\ }\href {\doibase https://doi.org/10.1016/j.ppnp.2022.104019}
  {\bibfield  {journal} {\bibinfo  {journal} {Progress in Particle and Nuclear
  Physics}\ }\textbf {\bibinfo {volume} {129}},\ \bibinfo {pages} {104019}
  (\bibinfo {year} {2023})}\BibitemShut {NoStop}%
\bibitem [{\citenamefont {Alvarez-Ruso}\ \emph {et~al.}(2025)\citenamefont
  {Alvarez-Ruso}, \citenamefont {Ankowski}, \citenamefont {Ashkenazi},
  \citenamefont {Barrow}, \citenamefont {Betancourt}, \citenamefont {Borah},
  \citenamefont {Athar}, \citenamefont {Catano-Mur}, \citenamefont {Coloma},
  \citenamefont {Dunne} \emph {et~al.}}]{Alvarez-Ruso:2025oak}%
  \BibitemOpen
  \bibfield  {author} {\bibinfo {author} {\bibfnamefont {L.}~\bibnamefont
  {Alvarez-Ruso}}, \bibinfo {author} {\bibfnamefont {A.~M.}\ \bibnamefont
  {Ankowski}}, \bibinfo {author} {\bibfnamefont {A.}~\bibnamefont {Ashkenazi}},
  \bibinfo {author} {\bibfnamefont {J.}~\bibnamefont {Barrow}}, \bibinfo
  {author} {\bibfnamefont {M.}~\bibnamefont {Betancourt}}, \bibinfo {author}
  {\bibfnamefont {K.}~\bibnamefont {Borah}}, \bibinfo {author} {\bibfnamefont
  {M.~S.}\ \bibnamefont {Athar}}, \bibinfo {author} {\bibfnamefont
  {E.}~\bibnamefont {Catano-Mur}}, \bibinfo {author} {\bibfnamefont
  {P.}~\bibnamefont {Coloma}}, \bibinfo {author} {\bibfnamefont
  {P.}~\bibnamefont {Dunne}},  \emph {et~al.},\ }\href
  {https://arxiv.org/abs/2503.23556} {\enquote {\bibinfo {title} {Neutrino
  scattering: Connections across theory and experiment},}\ } (\bibinfo {year}
  {2025}),\ \Eprint {http://arxiv.org/abs/2503.23556} {arXiv:2503.23556
  [hep-ex]} \BibitemShut {NoStop}%
\bibitem [{\citenamefont {Lozano}\ \emph {et~al.}(2025)\citenamefont {Lozano},
  \citenamefont {Silva}, \citenamefont {Caceres}, \citenamefont {Akhter},
  \citenamefont {Dar}, \citenamefont {Ansari}, \citenamefont {Ascencio},
  \citenamefont {Athar}, \citenamefont {Bonilla}, \citenamefont {Bravar} \emph
  {et~al.}}]{lozano:2025Minerva}%
  \BibitemOpen
  \bibfield  {author} {\bibinfo {author} {\bibfnamefont {A.}~\bibnamefont
  {Lozano}}, \bibinfo {author} {\bibfnamefont {G.}~\bibnamefont {Silva}},
  \bibinfo {author} {\bibfnamefont {G.}~\bibnamefont {Caceres}}, \bibinfo
  {author} {\bibfnamefont {S.}~\bibnamefont {Akhter}}, \bibinfo {author}
  {\bibfnamefont {Z.~A.}\ \bibnamefont {Dar}}, \bibinfo {author} {\bibfnamefont
  {V.}~\bibnamefont {Ansari}}, \bibinfo {author} {\bibfnamefont {M.~V.}\
  \bibnamefont {Ascencio}}, \bibinfo {author} {\bibfnamefont {M.~S.}\
  \bibnamefont {Athar}}, \bibinfo {author} {\bibfnamefont {J.~L.}\ \bibnamefont
  {Bonilla}}, \bibinfo {author} {\bibfnamefont {A.}~\bibnamefont {Bravar}},
  \emph {et~al.},\ }\href {https://arxiv.org/abs/2503.20043} {\enquote
  {\bibinfo {title} {Measurement of charged-current $\nu_\mu$ and
  $\bar{\nu}_\mu$ cross sections on hydrocarbon in a shallow inelastic
  scattering region},}\ } (\bibinfo {year} {2025}),\ \Eprint
  {http://arxiv.org/abs/2503.20043} {arXiv:2503.20043 [hep-ex]} \BibitemShut
  {NoStop}%
\bibitem [{\citenamefont {Acero}\ \emph {et~al.}(2023)\citenamefont {Acero},
  \citenamefont {Adamson}, \citenamefont {Aliaga}, \citenamefont {Anfimov},
  \citenamefont {Antoshkin}, \citenamefont {Arrieta-Diaz}, \citenamefont
  {Asquith}, \citenamefont {Aurisano}, \citenamefont {Back}, \citenamefont
  {Baird} \emph {et~al.}}]{Acero:2023Nova}%
  \BibitemOpen
  \bibfield  {author} {\bibinfo {author} {\bibfnamefont {M.~A.}\ \bibnamefont
  {Acero}}, \bibinfo {author} {\bibfnamefont {P.}~\bibnamefont {Adamson}},
  \bibinfo {author} {\bibfnamefont {L.}~\bibnamefont {Aliaga}}, \bibinfo
  {author} {\bibfnamefont {N.}~\bibnamefont {Anfimov}}, \bibinfo {author}
  {\bibfnamefont {A.}~\bibnamefont {Antoshkin}}, \bibinfo {author}
  {\bibfnamefont {E.}~\bibnamefont {Arrieta-Diaz}}, \bibinfo {author}
  {\bibfnamefont {L.}~\bibnamefont {Asquith}}, \bibinfo {author} {\bibfnamefont
  {A.}~\bibnamefont {Aurisano}}, \bibinfo {author} {\bibfnamefont
  {A.}~\bibnamefont {Back}}, \bibinfo {author} {\bibfnamefont {M.}~\bibnamefont
  {Baird}},  \emph {et~al.} (\bibinfo {collaboration} {NOvA Collaboration}),\
  }\href {\doibase 10.1103/PhysRevLett.130.051802} {\bibfield  {journal}
  {\bibinfo  {journal} {Phys. Rev. Lett.}\ }\textbf {\bibinfo {volume} {130}},\
  \bibinfo {pages} {051802} (\bibinfo {year} {2023})}\BibitemShut {NoStop}%
\bibitem [{\citenamefont {Lu}\ \emph {et~al.}(2021)\citenamefont {Lu},
  \citenamefont {Dar}, \citenamefont {Akbar}, \citenamefont {Andrade},
  \citenamefont {Ascencio}, \citenamefont {Barr}, \citenamefont {Bashyal},
  \citenamefont {Bellantoni}, \citenamefont {Bercellie}, \citenamefont
  {Betancourt} \emph {et~al.}}]{Lu:2021Minerva}%
  \BibitemOpen
  \bibfield  {author} {\bibinfo {author} {\bibfnamefont {X.-G.}\ \bibnamefont
  {Lu}}, \bibinfo {author} {\bibfnamefont {Z.~A.}\ \bibnamefont {Dar}},
  \bibinfo {author} {\bibfnamefont {F.}~\bibnamefont {Akbar}}, \bibinfo
  {author} {\bibfnamefont {D.}~\bibnamefont {Andrade}}, \bibinfo {author}
  {\bibfnamefont {M.}~\bibnamefont {Ascencio}}, \bibinfo {author}
  {\bibfnamefont {G.}~\bibnamefont {Barr}}, \bibinfo {author} {\bibfnamefont
  {A.}~\bibnamefont {Bashyal}}, \bibinfo {author} {\bibfnamefont
  {L.}~\bibnamefont {Bellantoni}}, \bibinfo {author} {\bibfnamefont
  {A.}~\bibnamefont {Bercellie}}, \bibinfo {author} {\bibfnamefont
  {M.}~\bibnamefont {Betancourt}},  \emph {et~al.},\ }\href {\doibase
  doi.org/10.1140/epjs/s11734-021-00296-6} {\bibfield  {journal} {\bibinfo
  {journal} {The European Physical Journal Special Topics}\ }\textbf {\bibinfo
  {volume} {230}},\ \bibinfo {pages} {4243} (\bibinfo {year}
  {2021})}\BibitemShut {NoStop}%
\end{thebibliography}

\end{document}